\documentclass[12pt]{article}
\begin{document}
\newcommand{\beq}{\begin{equation}}
\newcommand{\eeq}{\end{equation}}
\newcommand{\beqn}{\begin{eqnarray}}
\newcommand{\eeqn}{\end{eqnarray}}
\newcommand{\dpf}{\displaystyle\frac}
\newcommand{\no}{\nonumber}
\newcommand{\ep}{\epsilon}
\begin{center}
{\Large Friedmann equation and Cardy formula correspondence in brane universes}
\end{center}
\vspace{1ex}
\centerline{\large Bin
Wang$^{a,b,}$\footnote[1]{e-mail:binwang@fma.if.usp.br},
\ Elcio Abdalla$^{a,}$\footnote[2]{e-mail:eabdalla@fma.if.usp.br}
\ and Ru-Keng Su$^{c,b,}$\footnote[3]{e-mail:rksu@fudan.ac.cn}}
\begin{center}
{$^{a}$ Instituto De Fisica, Universidade De Sao Paulo, C.P.66.318, CEP
05315-970, \\ Sao Paulo, Brazil \\
$^{b}$ Department of Physics, Fudan University, Shanghai 200433,
P. R. China \\
$^{c}$ CCAST (World Lab), P.O.Box 8730, Beijing 100080, P. R. China}
\end{center}
\vspace{6ex}
\begin{abstract}
We study the brane with arbitrary tension $\sigma$ on the edge of various
black holes 
with AdS asymptotics. We investigate Friedmann equations governing the
motion of the 
brane universes and match the Friedmann equation to Cardy entropy formula.
\end{abstract}
\vspace{6ex} \hspace*{0mm} PACS number(s): 04.70.Dy, 98.80.Cq
\vfill
\newpage
Motivated by the well-known example of black hole entropy, an influential
holographic 
principle was put forward recently, which suggests that microscopic
degrees of freedom 
that build up the gravitational dynamics do reside not in the bulk
spacetime but on its 
boundary [1]. This principle has been viewed as a conceptual change in our
thinking about 
gravity. Maldacena's conjecture on AdS/CFT correspondence [2] is the first
example 
realizing such a principle. Subsequently, Witten [3] convincingly argued
that the 
entropy, energy and temperature of CFT at high temperatures can be
identified with the 
entropy, mass and Hawking temperature of the AdS black hole [4], which
further supports 
the holographic principle. In cosmological settings, testing the
holographic principle is 
somewhat subtle. Fischler and Susskind (FS) [5] 
have shown that for flat and open
Friedmann-Lemaitre-Robertson-Walker(FLRW) universes the area of the
particle horizon 
should bound the entropy on the backward-looking light cone. However
violation of FS 
bound was found for closed FLRW universes. Various different modifications
of the FS 
version
of the holographic principle have been raised subsequently [6-12]. In
addition to the 
study of holography in homogeneous cosmologies, attempts to generalize the
holographic 
principle to a generic realistic inhomogeneous cosmological setting were
carried out
in [13,14]. 

Recently, the very interesting study of the holographic principle in FLRW
universe 
filled
with CFT with a dual AdS description has been done by Verlinde [15]. He
revealed that 
when a universe-sized black hole can be formed, an interesting and surprising
correspondence appears between entropy of CFT and Friedmann equation
governing the 
radiation dominated closed FLRW universes. Generalizing Verlinde's
discussion to a 
broader class of universes including a cosmological constant [16], we
found that matching 
of Friedmann equation to Cardy formula holds for de Sitter closed and AdS flat
universes. 
However for the remaining de Sitter and AdS universes, the argument fails
due to breaking 
down of the general philosophy of the holographic principle. In high
dimensions, various 
other aspects of Verlinde's proposal have also been investigated in a
number of works 
[17].

In a very recent paper [18], further light on the correspondence between
Friedmann 
equation and Cardy formula has been shed from a Randall-Sundrum type
brane-world 
perspective [19]. Considering the CFT dominated universe as a co-dimension
one brane with 
fine-tuned tension in a background of an AdS black hole, Savonije and
Verlinde found the 
correspondence between Friedmann equation and Cardy formula for the
entropy of CFT when 
the brane crosses the black hole horizon. This result has been further
confirmed by 
studying a brane-universe filled with radiation and stiff-matter,
quantum-induced brane 
worlds and radially infalling brane [20]. The discovered relation between
Friedmann 
equation and Cardy formula for the entropy shed significant light on the
meaning  
of the holographic principle in a cosmological setting. However the
general proof for 
this correspondence for all CFTs is still difficult at the moment. It is
worthwhile to 
further check the validity of the correspondence in broader classes of
situations than 
[15,18]. This is the motivation of the present paper. In addition to
spherically 
symmetric AdS Schwarzschild black hole considered in the bulk background,
we will 
consider various black holes with AdS asymptotics including hyperbolic AdS
black holes 
and flat AdS black membrane. Instead of choosing a special value of the
brane tension to
tune the cosmological constant in the brane-universe to zero, in our
following study we 
will choose an arbitrary value of the tension on the brane to describe the
de Sitter and 
AdS brane universes. We will show that in de Sitter and AdS brane universes,
correspondence between the Friedmann equation and Cardy formula hold for
all values of 
curvature constants. The situation when a domain wall with matter is present in
addition to the 
background wall tension will also be addressed.

We start by considering a four-dimensional (4D) brane in a spacetime
described by a 
five-dimensional (5D) AdS black hole. The background metric is 
\beq  
ds^2_5=-fdt^2+f^{-1}dr^2+r^2d\Sigma^2_K,
\eeq
where
\beq  
f=k+\dpf{r^2}{L^2}-\dpf{m}{r^2}.
\eeq
$L$ is the curvature radius of AdS spacetime. $k$ takes the values $0, -1, +1$
corresponding to flat, open and closed geometrics, and $d\Sigma^2_k$ is
the corresponding 
metric on the unit three dimensional plane, hyperboloid or sphere. In the
case of $m=0$,
we have an exact 5D AdS space. For $m\not= 0$, the black hole horizon
locates at  
\beq           
r_H^2=\dpf{L^2}{2}(-k +\sqrt{k^2+4m/L^2}).
\eeq
The relation betwen $m$ to the Arnowitt-Deser-Misner (ADM) mass of the
hole $M$ is [21] 
\beq    
M=\dpf{3\omega_3}{16\pi G_5}m
\eeq
where $\omega_3$ is the volume of the unit 3 sphere.

As discussed in [18], we regard the brane as a boundary of background
AdS geometry. The location and the metric on the boundary become
time dependent. Choosing the gauge where 
\beq        
\dot{r}^2=f^2\dot{t}^2 -f,
\eeq
the metric on the brane is given by 
\beq       
ds^2_4= -d\tau^2+r^2(\tau)d\Sigma^2_3.
\eeq

Following [3,18], we learn that CFT lives on the brane, which is the
boundary of the AdS
hole. The energy for a CFT on a sphere with volume $V=r^3\omega_3$ is given by
$E=\dpf{L}{r}M$. The density of the CFT energy can be expressed as 
\beq       
\rho_{CFT}=E/V=\dpf{3mL}{16\pi G_5 r^4}.
\eeq

According to [3,22], the entropy of the CFT on the brane is equal to the
Bekenstein-Hawking entropy of the AdS black hole
\beq    
S_{CFT}(4D)=S_{BH}(5D)=\dpf{V_H}{4G_5}, \quad V_H=r_H^3\omega_3.
\eeq
The area of AdS equals to the volume of the corresponding spatial section
for an observer 
on the brane. The entropy density in the brane is 
\beq     
s=S/V=\dpf{r_H^3}{4G_5 r^3}=\dpf{r_H^3}{2G_4 L r^3}
\eeq
where $G_5=\dpf{G_4 L}{2}$.

The induced CFT temperature on the brane is 
\beq    
T=\dpf{L}{r}T_H,
\eeq
where $T_H$ is the Hawking temperature of the bulk black hole
\beq           
T_H=\dpf{1}{4\pi}f'(r_H)=\dpf{1}{2\pi}[\dpf{r_H}{L^2}+\dpf{m}{r_H^3}].
\eeq
The first law of thermodynamics expressed in terms of densities has the
form on the brane 
\beq     
Tds=d\rho_{CFT} + 3(\rho_{CFT} + P_{CFT} -Ts)\dpf{dr}{r}.
\eeq
Considering the equation of state $P_{CFT}=\rho_{CFT}/3$ and taking
$m=kr_H^2+r_H^4/L^2$, 
the term representing Casimir energy density derived from eqs(7,10,11) is 
\beq      
\dpf{3}{2}(\rho_{CFT}+P_{CFT}-Ts)=\dpf{\gamma}{r^2}
\eeq
where $\gamma=\dpf{3kr_H^2}{8\pi G_4 r^2}$.

With the given expressions for the entropy density $s$, CFT energy density
$\rho_{CFT}$ 
and $\gamma$, the entropy density can be expessed in the form of Cardy's
formula 
\beq  
s^2=(\dpf{4\pi}{3\sqrt{k}})^2\gamma (\rho_{CFT} - \dpf{\gamma}{r^2}).
\eeq
It is easy to check that this formula agrees to Eq.(30) in [18] regardless
of different values of $k$.

Now we start to study the motion of brane-universe and examine the
relation between the
Friedmann equation and Cardy formula in the brane-cosmology.

The equation of motion for the scalar factor $r(\tau)$ can be obtained
from Israel's 
matching conditions under the assumption of $z_2$ symmetry:
\beq      
K_{\mu\nu}=-8\pi G_5[T_{\mu\nu}-\dpf{1}{3}T^c_c\gamma_{\mu\nu}],
\eeq
where $T_{\mu\nu}$ is the energy momentum tensor on the brane and
$K_{\mu\nu}$ is the extrinsic curvature.

Introduce a stress-energy tensor on the brane [23]
\beq  
T_{ab}=-\sigma \gamma_{ab}
\eeq
where $\sigma$ is an arbitrary brane tension. From (15), the space
component of the 
junction condition gives 
\beq    
H^2=-\dpf{k}{r^2}+\dpf{m}{r^4}-\dpf{1-(\sigma/\sigma_c)^2}{L^2}
\eeq
where $\sigma_c=\dpf{3}{8\pi G_5 L}$ is the critical brane tension.
Taking $\sigma=\sigma_c$, (17) reduces to the Friedmann equation of CFT
radiation dominated brane universe without cosmological constant
discussed in [18]. If $\sigma>\sigma_c$ or $\sigma<\sigma_c$, the
brane-world is a de Sitter universe or AdS universe, respectively.

When the brane crosses the horizon $r_H$, eq(17) becomes
\beq 
H^2=\dpf{1}{L^2}(\dpf{\sigma}{\sigma_c})^2
\eeq
by taking $r=r_H$ and $m=kr_H^2+r_H^4/L^2$. The entropy density (9) can be
expressed at $r=r_H$ as 
\beq 
s=\dpf{H}{2G_4}(\sigma_c/\sigma).
\eeq
Putting (19) in (14) and taking $r=r_H$ and $m=kr_H^2+r_H^4/L^2$, (14)
exactly reproduces the first Friedmann equation.

Using the fact that $\dot{\rho}_{CFT}=-3H(P_{CFT}+\rho_{CFT})$ and
$P_{CFT}=\rho_{CFT}/3$, the second Friedmann equation can be obtained 
\beq 
\dot{H}=\dpf{k}{r^2}-\dpf{2m}{r^4}
\eeq
where (7) has been used. When the brane crosses the black hole
horizon, (20) can be expressed as
\beq 
\dot{H}=-\dpf{k}{r_H^2}-\dpf{2}{L^2}
\eeq
Employing (21) and (18), the CFT temperature at the black hole horizon is 
\beq 
T=-\dpf{\dot{H}}{2\pi H}(\dpf{\sigma}{\sigma_c}).
\eeq
From (13), we know 
\beq  
T=[\rho_{CFT}+P_{CFT}-\dpf{2\gamma}{3r^2}]/s.
\eeq
Substituting (22) and the values of $s, \gamma, \rho_{CFT}$ at the horizon, the
second Friedmann equation can be reproduced.

It is interesting to note that when the brane crosses the black hole
horizon, Friedmann equation coincides with Cardy formula. This result holds
independently of the value of the brane tension $\sigma$ and curvature
constant $k$. The correspondence between Friedmann equations and Cardy
formula is valid in all universes.

Now we consider the case in which motion of the brane universe results
from matter on the domain wall. We take the energy-momentum tensor
$T_{ab}=-\sigma \gamma_{ab}+\rho u_a u_b+P(\gamma_{ab}+u_a u_b)$, which
corresponds to matter with energy density $\rho$ and pressue $P$, in
addition to the background wall tension $\sigma$ [23]. For $\rho\ll
\sigma_c$, the angular components of (15) give 
\beq  
H^2=-\dpf{k}{r^2}+\dpf{m}{r^4}-[1-(\dpf{\sigma}{\sigma_c})^2-2(\dpf{\sigma}
{\sigma_c})\dpf{\rho}{\sigma_c}]/L^2
\eeq
when $\sigma=\sigma_c$, (24) reduces to the description in [23] on the
motion of brane universe. For $\sigma_c^2>\sigma^2 +2\sigma\rho$, (24) is
the Friedmann equation for AdS universe; while for $\sigma_c^2<\sigma^2
+2\sigma\rho$, (24) describes the de Sitter universe. The Friedmann
equation discussed in [18] is a specific case of (24) where
$\sigma=\sigma_c$ and $\rho=0$.

It has been argued that there is a possible duality relation between
classical gravity in 5D AdS bulk and a conformal field theory residing on
its boundary brane-world [3,22]. In this context, the second term in (24)
may be interpreted from the boundary theory point of view as the
contribution of a true radiation bath of the conformal fields whose
energy density satisfies (7).

Considering when the brane pass the black hole horizon, the Friedmann
equation (24) can be expressed as 
\beq    
H^2=[(\sigma/\sigma_c)^2+2(\sigma/\sigma_c)(\rho/\sigma_c)]/L^2=C/L^2.
\eeq
The CFT entropy density (9) can thus be written as 
\beq 
s=H/(2G_4\sqrt{C}).
\eeq
Substituting (26) into (14) and choosing values of $\gamma, \rho_{CFT}, r$
at the black hole horizon, it is easy to reproduce the first Friedmann
equation from the Cardy formula (14).

The second Friedmann equation can be obtained by differentiating (24) with
respect to $\tau$. For the radiation dominated universe with $P=\rho/3$,
\beq 
\dot{H}_1=\dpf{k}{r^2}-\dpf{2m}{r^4}-\dpf{4\rho\sigma}{L^2\sigma_c^2}.
\eeq
For matter dominated universe with $P=0$, the second Friedmann equation
reads 
\beq  
\dot{H}_2=\dpf{k}{r^2}-\dpf{2m}{r^4}-\dpf{3\rho\sigma}{L^2\sigma_c^2}.
\eeq
The CFT temperature (10) at the horizon may be expressed in the Hubble
constant and its time derivative as
\beq  
T_1=-\dpf{\dot{H}_1+\dpf{4\rho\sigma}{L^2\sigma_c^2}}{2\pi H}\sqrt{C} \quad 
{\rm for} \quad
P=\rho/3,
\eeq
\beq     
T_2= - \dpf{\dot{H}_2+\dpf{3\rho\sigma}{L^2\sigma_c^2}}{2\pi H}\sqrt{C}
\quad {\rm for} \quad  P=0.
\eeq
From (23), the second Friedmann equation can be reproduced at the black
hole horizon.

In summary, we have investigated the relationship between Friedmann
equation and Cardy formula in general cases of brane-universes. We found
that when the brane crosses the bulk AdS black hole horizon, the
correspondence between Friedmann equations and Cardy formula for CFT
entropy holds for all values of curvature constant $k$ on de Sitter and
AdS universes. At first sight, the result does not look consistent with that in
[16], where the match between Friedmann equation and Cardy
formula is obtained only for de Sitter closed and AdS flat universes. But we
claim that these two cases are different. Recall that in
deriving the result in [16], we need a universe-sized black hole to be
formed. However here in the moving domain-wall, we cannot define a
universe-sized black hole embedding into the moving brane-world [20]. The
generalized Cardy formula derived in [16] is the unification between
Bekenstein entropy bound and Hubble entropy bound. While the Cardy formula
obtained here exactly describes the entropy of radiation bath of conformal
fields. Our results generalized the discussion in [18] and shed further
light on the meaning of the holographic principle in cosmological setting.

ACKNOWLEDGEMENT: This work was partically supported by  
Fundac$\tilde{a}$o de Amparo $\grave{a}$ Pesquisa do Estado de
S$\tilde{a}$o Paulo (FAPESP) and Conselho Nacional de Desenvolvimento
Cient$\acute{i}$fico e Tecnol$\acute{o}$gico (CNPQ).  B. Wang would also  
like to acknowledge the support given by Shanghai Science and Technology  
Commission as well as NNSF, China under contract No. 10005004. R. K. Su's
work was 
supported by NNSF of China.

\end{document}